# Controlled catalyst transfer polymerization in graphene nanoribbon synthesis

Sai Ho Pun[1,†], Aidan Delgado[1,†], Christina Dadich[1], Adam Cronin[1], & Felix R. Fischer[1,2,3,4,*]

[1]Department of Chemistry, University of California, Berkeley, CA 94720, USA. [2]Materials Sciences Division, Lawrence Berkeley National Laboratory, Berkeley, CA 94720, USA. [3]Kavli Energy NanoScience Institute at the University of California Berkeley and the Lawrence Berkeley National Laboratory, Berkeley, California 94720, USA. [4]Bakar Institute of Digital Materials for the Planet, Division of Computing, Data Science, and Society, University of California, Berkeley, CA 94720, USA.

[†] These authors contributed equally to this work.

[*] Corresponding author

Exercising direct control over the unusual electronic structures arising from quantum confinement effects in graphene nanoribbons (GNRs) — atomically defined quasi one-dimensional (1D) strips of graphene — is intimately linked to geometric boundary conditions imposed by the bonding within the ribbon. Besides composition and position of substitutional dopant atoms, the symmetry of the unit cell, the width, length, and termination of a GNR are integral factors that collectively can give rise to highly tuneable semiconductors[1-3], innate metallicity arising from topological zero-mode engineering[4,5], or magnetic ordering in spin-polarized lattices[6]. Here we present a rational design that integrates each of these interdependent variables within a modular bottom-up synthesis. Our hybrid chemical approach relies on a catalyst transfer polymerization (CTP) that establishes uniform control over length, width, and end-groups. Complemented by a surface-assisted cyclodehydrogenation step, uniquely enabled by matrix-assisted direct (MAD) transfer protocols[7], geometry and functional handles encoded in a polymer template are faithfully mapped onto the structure of the corresponding GNR. Bond-resolved scanning tunnelling microscopy (BRSTM) and spectroscopy (STS) validate the robust correlation between polymer template design and GNR electronic structure and provide a universal and modular platform for the systematic exploration and seamless integration of functional GNRs with integrated circuit architectures.

Additive manufacturing approaches as realized in the chemical bottom-up synthesis of atomically defined GNRs from molecular building blocks in solution or on surfaces have dominated attempts to control integral structural parameters like crystallographic symmetry, width, and density/position of dopant atoms. While step-growth polymerizations[8] based on *Diels-Alder* cycloaddition[9-11] or transition metal catalysed *Suzuki-Miyaura*[12-15] and *Yamamoto*[16,17] cross-coupling reactions have advanced solution-based approaches, the on-surface synthesis of GNRs has largely relied on radical step-growth mechanisms[1-3,18]. Despite various examples, none of these techniques have demonstrated uncompromising, coextensive control over polymer sequencing underpinning the parameters that dictate the electronic structure of GNRs. While the design of the molecular building block used in solution- and surface-based polymerizations has customarily served to establish symmetry, width, and substitutional dopant levels, precise control over the length and functional termination of GNRs remains a veritable challenge and represent an indispensable stepping stone towards the successful integration of functional GNRs with photolithographically etched electronic circuit architectures and sensing nanotechnology.

Our strategy to exercise unprecedented length control and chemical functionalization of solution-synthesized GNR precursors builds on recent advancements in *Suzuki* catalyst transfer polymerization (SCTP) reactions[19-29]. Herein, a bifunctional monomer (**M**), e.g., the *o*-terphenylenes **1a,b** (Fig. 1a) featuring orthogonal functional groups, a halide (Br) and a boronic ester (Bpin) placed at the *para* positions of a central phenyl ring defining the axis of polymerization, is linearly extended into a one-dimensional (1D) polymer chain under the action of a palladium cross-coupling catalyst. A distinguishing feature of this class of polymerizations is that the Pd-catalyst, following the reductive elimination step, does not dissociate from the growing polymer chain. Instead, it remains coordinated to the π-system and migrates across the last aromatic ring in the polymer chain before reinserting into the C–Br bond priming the propagating catalyst for successive transmetallation with an arylboronate ester. We envisioned that a preformed aryl-halide Pd-complex (e.g., **2** in Fig. 1a) could serve as a competent initiator (**I**) and catalyst for SCTP of *o*-terphenylenes **1a,b**. The initiation step reliably transfers a functional end-group (e.g., the biphenyl highlighted in blue) to the inactive end of the growing polymer chain. In the absence of competing chain transfer ($k_{ct}$) and chain termination ($k_t$) steps, a fast initiation ($k_i$) and a near constant rate of propagation ($k_p$) ensures that the degree of polymerization (*DP*), and consequently the final length of the resulting polymers, can be controlled by the initial [**M**]$_0$/[**I**]$_0$ loading if $k_i > k_p \gg k_{ct} \sim k_t$. Once all monomer has been consumed the reactive chain ends can be terminated by coupling to a monofunctional terminator (**T**) (e.g., the phenylboronic acid **3** highlighted in orange) that caps the polymer chain and irreversibly dissociates the Pd catalyst. The resultant 1D polymers not only feature narrowly defined length distributions but two chemically distinct and individually addressable functional handles at either end of the polymer chain that map directly onto the structure of the corresponding *N* = 9 armchair GNRs (9-AGNRs, *N* is the number of C-atoms counted across the width of the ribbon).

A broad catalyst screening established the optimal conditions for a controlled living SCTP. The *o*-terphenylene monomer **1a** ($R^1$ = H) polymerizes readily under the action of a preformed Pd(RuPhos) aryl bromide catalyst **2**, in the presence of an aqueous base ($K_3PO_4$) in tetrahydrofuran (THF) at 24 °C to give the characteristic *p*-phenylene backbone of *poly*-**1a** (Fig. 1a). Size exclusion chromatography (SEC) calibrated to polystyrene (PS) standards reveals monomodal distributions with narrow dispersity (*Đ* = 1.19) at low [**M**]$_0$/[**I**]$_0$ < 20 that broaden at higher loadings as the solubility of the growing polymer chains decreases abruptly (Extended Data Fig. ED1). Polymers resulting from the reaction of *o*-terphenylene monomer **1b** ($R^1$ = $C_{12}H_{25}$) with **2** instead remain soluble in the reaction mixture at initial [**M**]$_0$/[**I**]$_0$ loadings of up to 100. Even after the consumption of all monomer the propagating catalyst species attached to the end of the growing polymer chain remains active and readily reacts with phenylboronic acid **3**, used here as a monofunctional chain-terminating agent. The resulting number-average molecular weight ($M_n$), dispersity (*Đ* = $M_w/M_n$; $M_w$ is the weight average molecular weight), degree of polymerization (*DP*), average polymer length ($\bar{l}$), termination efficiency, and isolated polymer yield at various [**M**]$_0$/[**I**]$_0$ loadings are summarized in Fig. 1b. SEC traces of crude *poly*-**1b** samples prepared by SCTP show monomodal distributions (Fig. 1c). The $M_n$ increases linearly with the initial [**M**]$_0$/[**I**]$_0$ loading from $M_n$ = 7.8 × 10$^3$ g

mol$^{-1}$ ([M]$_0$/[I]$_0$ = 20) to $M_n$ = 52.9 × 10$^3$ g mol$^{-1}$ ([M]$_0$/[I]$_0$ = 100) while the dispersity remains low ($Đ ≤$ 1.2) (Fig. 1d). SEC samples of *poly*-**1b** taken at different timepoints from a SCTP reaction ([M]$_0$/[I]$_0$ = 25) reveal a linear correlation between $M_n$ and the monomer conversion indicative of a living chain-growth mechanism (Fig. 1e). MALDI-TOF mass spectrometry of *poly*-**1b** samples shows a characteristic family of peaks separated by the monomer repeat unit Δ$m/z$ = 564.8 ± 2.8 (Fig. 1f). The mass of detected molecular ions [M]$^+$ = [*poly*-**1b**–CHO], correspond to the terminated polymers and spans from 5 < DP < 24 for [M]$_0$/[I]$_0$ = 20. Despite the loss of the labile aldehyde group during ionization, MALDI-TOF supports the formation of telechelic polymers featuring both chemically unique end-groups transferred as part of initiation and termination steps. $^1$H NMR spectra of *poly*-**1b** show two characteristic resonance peaks at $δ$ = 3.27 ppm and $δ$ = 9.88 ppm corresponding to the dimethyl acetal and the aldehyde group, respectively (Fig. 1g). Acid catalysed deprotection of the acetal in *poly*-**1b** (Fig. 1a) leads to the loss of signal at $δ$ = 3.27 ppm along with the emergence of a new resonance at $δ$ = 9.95 ppm corresponding to the second free aldehyde in *poly*-**4b** transferred as part of the initiation step. Integration of the corresponding $^1$H NMR resonances for *poly*-**1b** samples prepared from [M]$_0$/[I]$_0$ loadings of 20 < [M]$_0$/[I]$_0$ < 100 suggest termination efficiencies in excess of 90% (Fig. 1b).

We next explored the conversion of solubilized, end-functionalized, and length-controlled samples of *poly*-**4b** into the corresponding 9-AGNRs (Fig. 2a). *Scholl*-type[30,31] cyclodehydrogenation reactions mediated by either FeCl$_3$ or 2,3-dichloro-5,6-dicyano-1,4-benzoquinone (DDQ)/trifluoromethanesulfonic acid (TfOH) as oxidants yielded amorphous precipitates commonly associated with the formation of GNRs. UV-Vis spectroscopy of sparingly soluble 9-AGNR-**4b** prepared from *poly*-**4b** ($M_n$ = 7.8 × 10$^3$ g mol$^{-1}$) show a bathochromic shift and broadening of the absorption spectrum indicative of the transition from a *poly*-(*p*-phenylene) chromophore ($λ_{max}$ = 292 nm) to the extended polycyclic aromatic backbone in 9-AGNRs ($λ_{max}$ = 293, 342, 482, 530, 768 nm) (Fig. 2b)[32]. Vibrational (IR) spectroscopy recorded on both *poly*-**4b** precursors and the corresponding 9-AGNRs shows the characteristic alky and aryl C–H stretching and bending modes (2918 cm$^{-1}$, 2848 cm$^{-1}$, and 1456 cm$^{-1}$, respectively) along with the aldehyde C=O stretching mode a 1714 cm$^{-1}$. An increase in the aromatic C=C stretching mode at 1557 cm$^{-1}$ can be attributed to the expansion of the polycyclic aromatic backbone in 9-AGNRs. Raman spectroscopy reveals the characteristic signature of D and G peaks, 1320 cm$^{-1}$ and 1598 cm$^{-1}$ respectively, along with 2D, D+G and 2G overtones (Fig. 2b). The radial breathing-like mode (RBLM) commonly observed in samples of on-surface grown 9-AGNRs (RBLM ~ 310–320 cm$^{-1}$) is absent from the spectrum or too weak to be detected over the signal background[32].

The chemical structure of solution synthesized 9-AGNRs, the sparingly soluble products of a *Scholl*-type oxidation of *poly*-**4b**, was further characterized using cryogenic (4 K) scanning tunnelling microscopy (STM) in ultra-high vacuum (UHV). Samples of 9-AGNR-**4b** dispersed in an inert matrix of pyrene were deposited on Au(111) surfaces using matrix assisted direct (MAD) transfer techniques[7]. Topographic STM images (Extended Data Fig. ED2) recorded on samples annealed at 353 K for 12 h to induce lateral diffusion and traceless sublimation of the pyrene matrix, show irregular graphitic structures (apparent height 0.20 nm ± 0.05 nm) that cannot be mapped onto the crystallographic unit cell of the 9-AGNR backbone. The obvious dissonance between ensemble-based characterization, i.e., NMR, mass spectrometry, Raman, IR, UV/Vis, and single molecule resolved STM imaging uniquely enabled by MAD transfer techniques echoes the structural complexity emerging from radical cation rearrangements and fragmentations intrinsic to a given PAH scaffold when subjected to *Scholl*-type reaction conditions. The advent of MAD transfer in combination with bond-resolved STM imaging tools has irreversibly raised the burden of evidence associated with the structural characterization of solution synthesized GNR approaches.

To overcome the technical challenge associated with supressing undesired side-reactions during the *Scholl*-type oxidation of *poly*-**4a** we adopted a hybrid GNR growth process. While SCTP provides unprecedented control over and ready access to narrow dispersity, end-functionalized 9-AGNR polymer precursors, MAD transfer followed by surface-assisted cyclodehydrogenation under UHV conditions ensures the formation of only the desired C–C bonds that map onto the unit cell of the fused GNR backbone. Guided by this idea we prepared high-dilution samples of low dispersity *poly*-**4a** ([**M**]$_0$/[**I**]$_0$ = 20) in a pyrene matrix for MAD transfer onto a clean Au(111) surface held at 297 K (see detailed procedure in Methods section). Molecule-decorated surfaces were annealed in UHV for 12 h at 353 K to induce the sublimation of the bulk pyrene matrix and diffusion of *poly*-**4a** across the surface. Further annealing at 648 K for 30 min induces a thermal cyclodehydrogenation that leads to the fully fused 9-AGNR backbone. Figure 2e shows a constant-current STM image of a sub-monolayer coverage of 9-AGNR-**4a** on Au(111) at $T$ = 4 K. Large-area topographic scans reveal 9-AGNR-**4a** featuring narrow length distributions ($\bar{l}$ = 7.9 ± 1.3 nm) commensurate with the average *DP* of *poly*-**4a** derived from SEC (Fig. 1b). 9-AGNRs assemble into long range ordered domains that align along the elbows of the herringbone reconstruction of the Au(111) surface. The average lateral spacing between GNRs is > 3 nm precluding any direct interaction between individual ribbons. Higher resolution constant-current STM images (Fig. 2f, Extended Data Fig. ED3) reveal GNRs featuring atomically smooth armchair edges with an apparent height and width of 0.19 nm ± 0.01 nm and 1.00 nm ± 0.01 nm, respectively, consistent with the formation of the fully conjugated 9-AGNR backbone. A majority of ribbons feature at least one (> 90%) or both (> 70%) distinctive end-groups, i.e. the 2-bromo-1,1'-biphenyl and the phenylboronic acid (blue and orange phenyl rings in Fig. 1a and 2a), transferred as part of the SCTP initiation and termination steps (we will herein use these groups to refer to the initiator-end and the terminator-end of the 9-AGNR, respectively). Singular indentations along the armchair edges of 9-AGNRs result from a commonly observed phenyl deletion attributed to the occasional cleavage of C–C bonds during the cyclodehydrogenation step (Extended Data Fig. ED4b). Tip functionalized bond-resolved STM (BRSTM) images recorded on a representative fully cyclized 9-AGNR-**4a** (*DP* = 15) (Fig. 2g) and a shorter ribbon (*DP* = 11) (Fig. 2h) show both end-groups transferred during the initiation and termination step. While BRSTM images reveal the characteristic pattern of four laterally fused benzene rings (i.e. a tetracene core) spanning the width of the 9-AGNR backbone closer inspection reveals that the thermally labile aldehyde groups lining the ends of *poly*-**4a**, used as diagnostic $^1$H NMR markers during the SCTP (Fig. 1g), are cleaved during the cyclodehydrogenation step. This observation is consistent with MALDI-TOF experiments that only show the mass of the molecular ions [M]$^+$ = [*poly*-**1b**–CHO] (Fig. 1f). The robust correlation between solution-based SCTP polymer characterization and the length distribution derived from STM of 9-AGNRs resulting from on-surface cyclodehydrogenation highlights the unrivalled structural control innate to our hybrid SCTP/MAD transfer bottom-up approach.

We herein used differential conductance spectroscopy to explore the local electronic structure of 9-AGNR-**4a** emerging from SCTP polymers. Typical d$I$/d$V$ point spectra recorded along the armchair edge of a pristine 9-AGNR-**4a** (Fig. 3a, positions marked with red crosses in BRSTM image inset) show five prominent features. A broad peak centred at $V_s$ = –0.25 ± 0.05 V (*Peak* 2) and a sharp signature at $V_s$ = +1.10 ± 0.05 V (*Peak* 3) have previously been assigned to the valence (VB) and conduction band (CB) edge, respectively, giving rise to a semiconducting gap $E_g$ ~ 1.4 eV on Au(111)[33-35]. The peaks at $V_s$ = –1.30 ± 0.05 V (*Peak* 1), $V_s$ = 1.50 ± 0.05 V (*Peak* 4), and $V_s$ = 1.90 ± 0.05 V (*Peak* 5) correspond to the VB–1, CB+1, and CB+2 band edges, respectively. d$I$/d$V$ point spectra recorded near either the initiator (position marked with blue crosses in BRSTM image inset) or the terminator end of the ribbon (position marked with orange crosses in BRSTM image inset) mirror the bulk VB and CB states but feature three additional low-lying end-states (ES) at $V_s$ = –0.50 ± 0.05 V (ES 1), $V_s$ = –0.60 ± 0.05 V (ES 2), and $V_s$ = –0.90 ± 0.05 V (ES 3) not observed in 9-AGNRs fabricated through on-surface polymerization and cyclodehydrogenation[33-35]. Differential conductance maps recorded at biases corresponding to the *Peaks* 3–5 in the point spectra show the characteristic nodal patterns associated with the CB+2, CB+1, and CB lining the armchair edges of the ribbon (Figs. 3b–d). At negative bias across the range of –0.10 V > $V_s$ > –0.30 V (Figs. 3e–g) d$I$/d$V$ maps are saturated by the Au(111) surface state. Despite the large background signal Fig. 3f shows the distinctive signature of the VB (*Peak* 2) — bright lobes lining the armchair edges contrasted by a dark featureless GNR backbone. The same nodal pattern is echoed in a map corresponding to the VB–1 (*Peak* 1) edge (Fig. 3k). Unique to 9-AGNRs synthesized through our hybrid SCTP/MAD transfer approach are the characteristic ES associated with the functional end-groups transferred during the initiation and termination steps of the polymerization. A differential conductance map recorded at $V_s$ = –0.50 V (ES 1) shows three bright lobes lining the lower (initiator) end of the ribbon in Fig. 3h while contrast along the remainder of the GNR is subdued (residual signal along the armchair edges represents contributions from the VB). Likewise, d$I$/d$V$ mapping at a bias of $V_s$ = –0.60 V (ES 2) shows two bright spots flanking the cove-type structure lining the upper (terminator) end of the GNR (Fig. 3i). Both ends of the ribbon are prominently featured in differential conductance maps recorded at $V_s$ = –0.90 V (ES 3) (Fig. 3j). The correspondence between the chemical structure of initiator- and terminator-ends derived from BRSTM and the emergence of distinctive peaks and nodal patterns at discrete energies in differential conductance spectroscopy is compelling evidence for highly localized end-states and implies that targeted chemical design of SCTP initiator and terminator groups seamlessly translates into the functional engineering of electronic end-states in bottom-up synthesized GNRs. Our hybrid SCTP/MAD transfer process thus not only provides exquisite control over the length, width and end-groups of bottom-up synthesized GNRs but presents an unparalleled tool to access and fine-tune the band alignment and contact resistance at the critical metal-semiconductor interface of nanoribbon device architectures, paving the way for technological advancements in the field of carbon-based nanoelectronics sensing and computing.

**References**

1   Cai, J. M. *et al.* Atomically precise bottom-up fabrication of graphene nanoribbons. *Nature* **466**, 470-473, (2010).

2   Rizzo, D. J. *et al.* Topological band engineering of graphene nanoribbons. *Nature* **560**, 204-208, (2018).

3   Groning, O. *et al.* Engineering of robust topological quantum phases in graphene nanoribbons. *Nature* **560**, 209-213, (2018).

4   Rizzo, D. J. *et al.* Inducing metallicity in graphene nanoribbons via zero-mode superlattices. *Science* **369**, 1597-1603, (2020).

5   McCurdy, R. D. *et al.* Robust Metallic Zero-Mode States in Olympicene Graphene Nanoribbons. *arXiv*, 2302.08446, (2023).

6   Blackwell, R. E. *et al.* Spin splitting of dopant edge state in magnetic zigzag graphene nanoribbons. *Nature* **600**, 647-652, (2021).

7   McCurdy, R. D. *et al.* Synergetic Bottom-Up Synthesis of Graphene Nanoribbons by Matrix-Assisted Direct Transfer. *J. Am. Chem. Soc.* **143**, 4174-4178, (2021).

8   Yoon, K. Y. & Dong, G. B. Liquid-phase bottom-up synthesis of graphene nanoribbons. *Mater. Chem. Front.* **4**, 29-45, (2020).

9   Wu, J. S. *et al.* From branched polyphenylenes to graphite ribbons. *Macromolecules* **36**, 7082-7089, (2003).

10  Narita, A. *et al.* Synthesis of structurally well-defined and liquid-phase-processable graphene nanoribbons. *Nat. Chem.* **6**, 126-132, (2014).

11  Narita, A. *et al.* Bottom-Up Synthesis of Liquid-Phase-Processable Graphene Nanoribbons with Near-Infrared Absorption. *ACS Nano* **8**, 11622-11630, (2014).

12  Yang, X. Y. *et al.* Two-dimensional graphene nanoribbons. *J. Am. Chem. Soc.* **130**, 4216-4217, (2008).

13  Schwab, M. G. *et al.* Bottom-Up Synthesis of Necklace-Like Graphene Nanoribbons. *Chem. Asian J.* **10**, 2134-2138, (2015).

14  Li, G., Yoon, K. Y., Zhong, X. J., Zhu, X. Y. & Dong, G. B. Efficient Bottom-Up Preparation of Graphene Nanoribbons by Mild Suzuki-Miyaura Polymerization of Simple Triaryl Monomers. *Chem.-Eur. J.* **22**, 9116-9120, (2016).

15  Li, G. *et al.* A modular synthetic approach for band-gap engineering of armchair graphene nanoribbons. *Nat. Comm.* **9**, 1687, (2018).

16  Schwab, M. G. *et al.* Structurally Defined Graphene Nanoribbons with High Lateral Extension. *J. Am. Chem. Soc.* **134**, 18169-18172, (2012).

17  Jansch, D. *et al.* Ultra-Narrow Low-Bandgap Graphene Nanoribbons from Bromoperylenes – Synthesis and Terahertz-Spectroscopy. *Chem.-Eur. J.* **23**, 4870-4875, (2017).

18  Ruffieux, P. *et al.* On-surface synthesis of graphene nanoribbons with zigzag edge topology. *Nature* **531**, 489-492, (2016).

19  Yokoyama, A. *et al.* Chain-growth polymerization for the synthesis of polyfluorene via Suzuki-Miyaura coupling reaction from an externally added initiator unit. *J. Am. Chem. Soc.* **129**, 7236-7237, (2007).

20  Yokozawa, T. & Yokoyama, A. Chain-Growth Condensation Polymerization for the Synthesis of Well-Defined Condensation Polymers and π-Conjugated Polymers. *Chem. Rev.* **109**, 5595-5619, (2009).

21  Elmalem, E., Kiriy, A. & Huck, W. T. S. Chain-Growth Suzuki Polymerization of n-Type Fluorene Copolymers. *Macromolecules* **44**, 9057-9061, (2011).

22  Zhang, H. H., Xing, C. H. & Hu, Q. S. Controlled Pd(0)/*t*-Bu$_3$P-Catalyzed Suzuki Cross-Coupling Polymerization of AB-Type Monomers with PhPd(*t*-Bu$_3$P)I or Pd$_2$(dba)$_3$/*t*-Bu$_3$P/ArI as the Initiator. *J. Am. Chem. Soc.* **134**, 13156-13159, (2012).

23  Elmalem, E., Biedermann, F., Johnson, K., Friend, R. H. & Huck, W. T. S. Synthesis and Photophysics of Fully pi-Conjugated Heterobis-Functionalized Polymeric Molecular Wires via Suzuki Chain-Growth Polymerization. *J. Am. Chem. Soc.* **134**, 17769-17777, (2012).

24  Zhang, H. H., Xing, C. H., Hu, Q. S. & Hong, K. L. Controlled Pd(0)/*t*-Bu$_3$P-Catalyzed Suzuki Cross-Coupling Polymerization of AB-Type Monomers with ArPd(*t*-Bu$_3$P)X or Pd$_2$(dba)$_3$/*t*-Bu$_3$P/ArX as the Initiator. *Macromolecules* **48**, 967-978, (2015).

25  Zhang, H. H., Hu, Q. S. & Hong, K. Accessing conjugated polymers with precisely controlled heterobisfunctional chain ends via post-polymerization modification of the OTf group and controlled Pd(0)/*t*-Bu$_3$P-catalyzed Suzuki cross-coupling polymerization. *Chem. Commun.* **51**, 14869-14872, (2015).

26  Dong, J., Guo, H. & Hu, Q. S. Controlled Pd(0)/Ad$_3$P-Catalyzed Suzuki Cross-Coupling Polymerization of AB-Type Monomers with Ad$_3$P-Coordinated Acetanilide-Based Palladacycle Complex as Initiator. *ACS Macro Lett.* **6**, 1301-1304, (2017).

27  Seo, K. B., Lee, I. H., Lee, J., Choi, I. & Choi, T. L. A Rational Design of Highly Controlled Suzuki-Miyaura Catalyst-Transfer Polycondensation for Precision Synthesis of Polythiophenes and Their Block Copolymers: Marriage of Palladacycle Precatalysts with MIDA-Boronates. *J. Am. Chem. Soc.* **140**, 4335-4343, (2018).

28  Lee, J. *et al.* Universal Suzuki-Miyaura Catalyst-Transfer Polymerization for Precision Synthesis of Strong Donor/Acceptor-Based Conjugated Polymers and Their Sequence Engineering. *J. Am. Chem. Soc.* **143**, 11180-11190, (2021).

29  Kim, H., Lee, J., Kim, T., Cho, M. & Choi, T. L. Precision Synthesis of Various Low-Bandgap Donor-Acceptor Alternating Conjugated Polymers via Living Suzuki-Miyaura Catalyst-Transfer Polymerization. *Angew. Chem. Int. Ed.* **61**, e202205828 (2022).

30  Grzybowski, M., Skonieczny, K., Butenschon, H. & Gryko, D. T. Comparison of Oxidative Aromatic Coupling and the Scholl Reaction. *Angew. Chem. Int. Ed.* **52**, 9900-9930, (2013).

31  Zhang, Y. Q., Pun, S. H. & Miao, Q. The Scholl Reaction as a Powerful Tool for Synthesis of Curved Polycyclic Aromatics. *Chemical Reviews* **122**, 14554-14593, (2022).

32  Barin, G. B. *et al.* Surface-Synthesized Graphene Nanoribbons for Room Temperature Switching Devices: Substrate Transfer and ex Situ Characterization. *ACS Appl. Nano Mater.* **2**, 2184-2192, (2019).

33  Talirz, L. *et al.* On-Surface Synthesis and Characterization of 9-Atom Wide Armchair Graphene Nanoribbons. *ACS Nano* **11**, 1380-1388, (2017).

34  Merino-Diez, N. *et al.* Width-Dependent Band Gap in Armchair Graphene Nanoribbons Reveals Fermi Level Pinning on Au(111). *ACS Nano* **11**, 11661-11668, (2017).

35  Rizzo, D. J. *et al.* Rationally Designed Topological Quantum Dots in Bottom-Up Graphene Nanoribbons. *ACS Nano* **15**, 20633-20642, (2021).

**Acknowledgements**

This work was primarily funded by the U.S. Department of Energy (DOE), Office of Science, Basic Energy Sciences (BES), Materials Sciences and Engineering Division under contract no. DE-AC02-05-CH11231 (Nanomachine program KC1203) (molecular design, catalyst development, and polymer synthesis) and contract no. DE-SC0023105 (surface growth). STM characterization was supported by the U.S. Department of Defense (DOD), Office of Naval Research (ONR) under award no. N00014-19-1-2503 and STS analysis by the National Science Foundation (NSF) under award no. CHE-2203911. Part of this research program was generously supported by the Heising-Simons Faculty Fellows Program at UC Berkeley. STM instruments are supported in part by the U.S. Department of Defense (DOD), Office of Naval Research (ONR) under award no. N00014-20-1-2824. The authors thank the College of Chemistry (CoC) for use of resources at their NMR and small molecule X-ray crystallography facility. The authors thank their staff Hasan Çelik and Nicholas Settineri for assistance. Instruments in CoC-NMR facility are supported in part by NIH S10-OD024998. Instruments in CoC-X-ray facility are supported in part by NIH S10-RR027172.

**Author Contributions**

S.H.P., A.D., and F.R.F. initiated and conceived the research, S.H.P. and C.D. designed, synthesized, and characterized monomer, catalyst, and polymer, A.D and A.C. performed on-surface synthesis, STM characterization and data analysis. S.H.P, C.D., A.D. and F.R.F. wrote the manuscript. All authors contributed to the scientific discussion.

**Author Information**

The authors declare no competing financial interests. Correspondence and requests for materials should be addressed to ffischer@berkeley.edu.

**Main Figures and Legends**

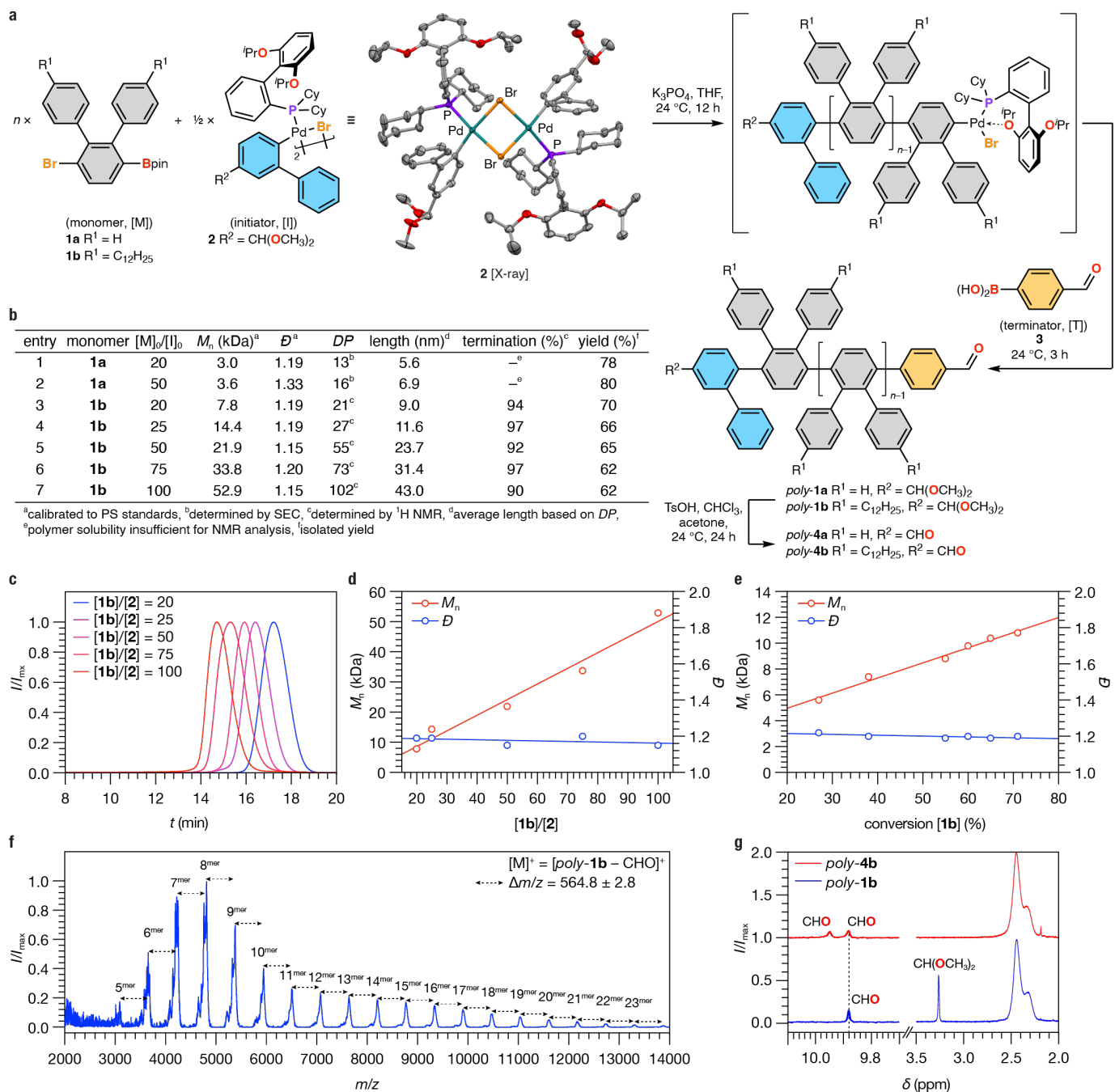

**Figure 1 | *Suzuki* catalyst transfer polymerization. a,** Schematic representation of the living SCTP of *o*-terphenylene monomers **1a,b** [**M**] initiated by a Pd(RuPhos) aryl bromide complex **2** [**I**]. The reaction is terminated by a monofunctional boronic acid **3** [**T**]. Single X-ray crystal structure of **2**. Thermal ellipsoids are drawn at the 50% probability level. Colour coding: C (gray), O (red), P (purple), Br (orange), Pd

(turquoise). H-atoms are omitted for clarity. **b,** Summary of SEC data, estimated length, termination efficiency, and isolated yields of *poly*-**1a,b** as a function of initial loading $[M]_0/[I]_0$. **c,** SEC traces of *poly*-**1b** samples obtained from various initial $[M]_0/[I]_0$ loadings. **d,** Linear correlation between $M_n$ and $[M]_0/[I]_0$. Đ remains constant for different $[M]_0/[I]_0$. **e,** Evolution of $M_n$ and Đ as a function of monomer conversion. **f,** MALDI of *poly*-**1b** ($[M]_0/[I]_0 = 20$) showing a family of polymers separated by the mass of the monomer unit $\Delta m/z$. The molecular ion peaks correspond to the mass of the polymer minus the labile aldehyde group. **g,** $^1$H-NMR of *poly*-**1b** (blue) and *poly*-**4b** (red). Deprotection of the dimethyl acetal ($\delta = 3.27$ ppm) leads to the emergence of a resonance at $\delta = 9.95$ ppm associated with the second aldehyde group.

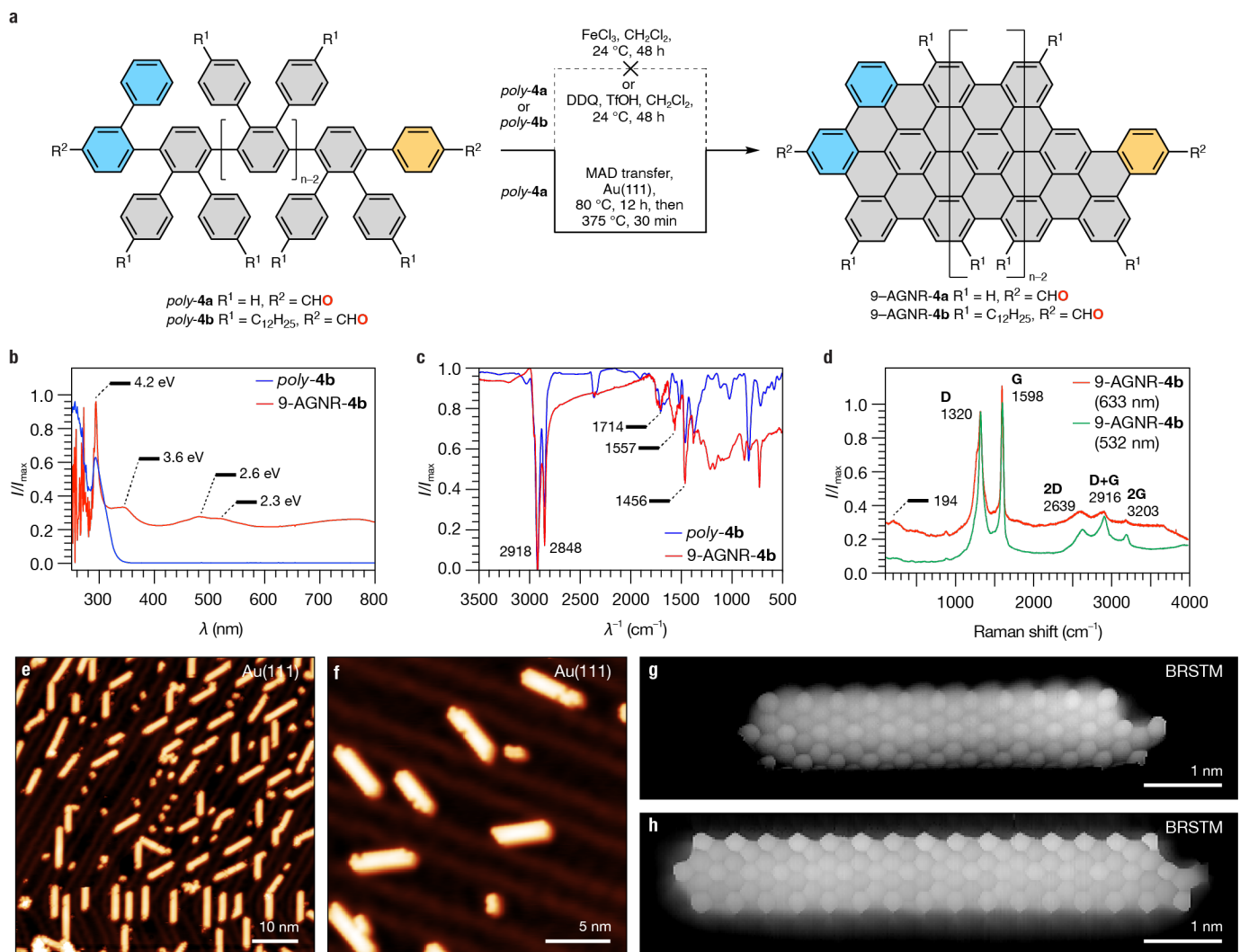

**Figure 2 | Solution based cyclodehydrogenation and hybrid surface assisted synthesis of 9-AGNRs from polymer precursors. a,** Schematic representation of the solution-based *Scholl*-type reaction of *poly*-**4a** and *poly*-**4b** (top) and surface-assisted cyclodehydrogenation of *poly*-**4a** on Au(111) (bottom). **b,** UV-Vis spectra of *poly*-**4b** and the resulting GNR products following solution-based oxidative cyclodehydrogenation (*Scholl*-type reaction). **c,** IR spectra of *poly*-**4b** and the resulting GNR products following solution-based oxidative cyclodehydrogenation (*Scholl*-type reaction). **d,** Raman spectrum of GNR products resulting from solution-based oxidative cyclodehydrogenation (*Scholl*-type reaction) of *poly*-**4b**. Excitation laser: 633 and 532 nm (spectra are offset for clarity). **e,** Large area topographic STM image (constant current) of a MAD transferred annealed sample of *poly*-**4a** ($[M]_0/[I]_0 = 20$) on Au(111) ($V_s = 100$ mV, $I_t = 20$ pA). **f,** Topographic STM image (constant current) of a MAD transferred annealed sample

of *poly*-**4a** ($[M]_0/[I]_0 = 20$) on Au(111) ($V_s = 50$ mV, $I_t = 20$ pA). **g–h,** BRSTM images (constant height) of a pristine 9-AGNR-**4a** (*DP* = 11 and 15) grown on Au(111) from *poly*-**4a** ($[M]_0/[I]_0 = 20$) using the hybrid SCTP/surface-assisted cyclodehydrogenation approach ($V_s = 10$ mV, $V_{ac} = 10$ mV, $f = 455$ Hz, CO-functionalized tip). All STM experiments performed at $T = 4$ K.

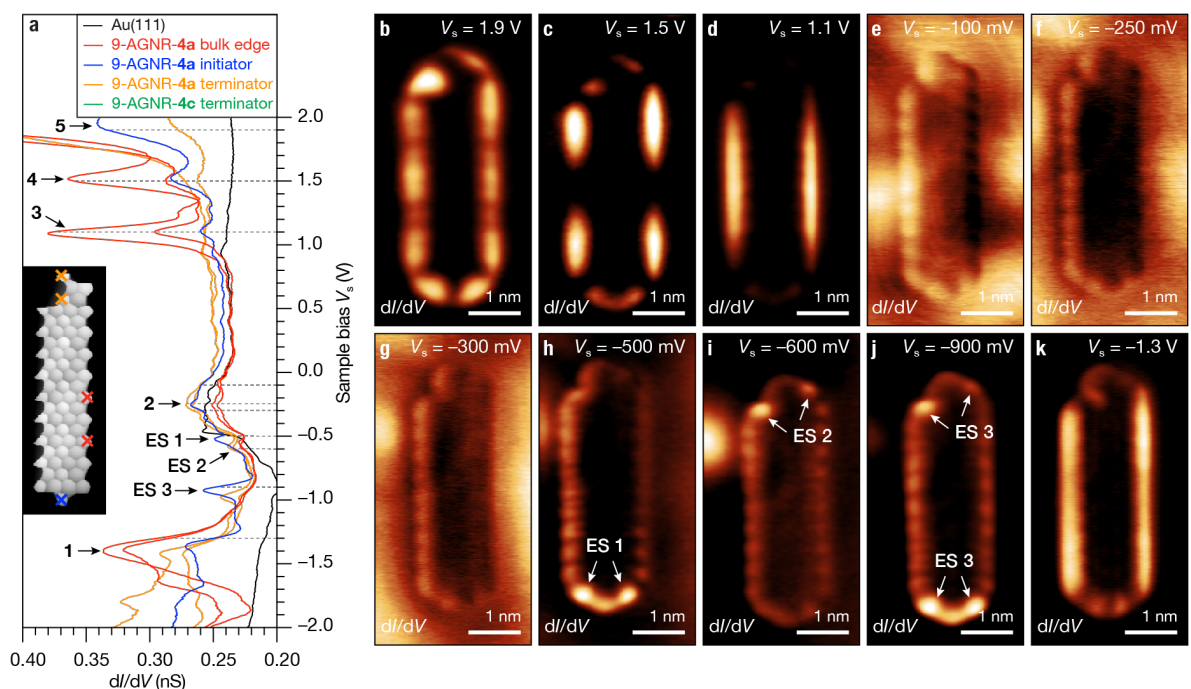

**Figure 3 | Electronic structure and differential conductance maps of end-functionalized 9-AGNRs. a,** d$I$/d$V$ point spectra of a 9-AGNR-**4a** on Au(111) recorded at the position marked in the BRSTM inset (Au(111) surface state, black; armchair edge, red; initiator end, blue; terminator end, orange; $V_{ac}$ = 10 mV, $f$ = 455 Hz, CO-functionalized tip). **b–k,** Differential conductance maps (d$I$/d$V$) recorded at the highlighted bias voltages ($V_s$) on a 9-AGNR-**4a** ($DP$ = 9) shown in the inset of **a**. Arrows highlight the localization of electronic end-states (ES 1, ES 2, ES 3) introduced as part of the SCTP.

**Data Availability**

All data presented in the main text and the supplementary materials are available from the corresponding authors upon reasonable request.

**Methods**

**Precursor Synthesis and GNR Growth.** Full details of the synthesis and characterization of **1a,b**, **2**, as well as polymers *poly*-**1a,b** and *poly*-**4a,b** are given in the Supplementary Information. 9-AGNRs were grown from *poly*-**4a** on Au(111)/mica films under UHV conditions. Atomically clean Au(111) surfaces were prepared through iterative $Ar^+$ sputter/anneal cycles. Sub-monolayer coverage of *poly*-**4a** on atomically clean Au(111) was obtained by MAD transfer. *poly*-**4a** was dispersed in molten pyrene (pyrene:*poly*-**4a** = 5000:1 w/w) and flash frozen at 77 K. The glassy solid was ground to a fine powder and sprinkled onto an atomically clean Au(111) substrate in ambient air at $T$ = 298 K. The sample was transferred into UHV ($p < 10^{-9}$ Pa) and annealed at $T$ = 353 K for $t$ ~12 h to induce the slow sublimation of pyrene and diffusion of polymers across the surface. The surface temperature was slowly ramped ($\leq 2$ K $min^{-1}$) to 648 K and held at this temperature for 30 min to induce the radical-step growth polymerization and on-surface cyclodehydrogenation. MAD transfer samples of 9-AGNR-**4b** and 9-AGNR-**4b** resulting from *Scholl*-type oxidative cyclodehydrogenation in solution were prepared following the same procedure.

**STM Measurements.** All STM experiments were performed using a commercial OMICRON LT-STM held at $T$ = 4 K using PtIr STM tips. STM tips were optimized for scanning tunnelling spectroscopy using an automated tip conditioning program[36]. d$I$/d$V$ measurements were recorded with CO-functionalized STM tips using a lock-in amplifier with a modulation frequency of 455 Hz and a modulation amplitude of $V_{RMS}$ = 11 mV. d$I$/d$V$ point spectra were recorded under open feedback loop conditions. d$I$/d$V$ maps were collected under constant current conditions. BRSTM images were obtained by mapping the out-of-phase d$I$/d$V$ signal collected during a constant-height d$I$/d$V$ map at 10 mV. Peak positions in d$I$/d$V$ point spectroscopy were determined by fitting the spectra with Lorentzian peaks. Each peak position is based on an average of ~10 spectra collected on various GNRs with different tips, all of which were first calibrated to the Au(111) Shockley surface state.

**Methods References**

36  Wang, S. K., Zhu, J. M., Blackwell, R. & Fischer, F. R. Automated Tip Conditioning for Scanning Tunnelling Spectroscopy. *J. Phys. Chem. A* **125**, 1384-1390, (2021).

**Extended Data Figures and Legends**

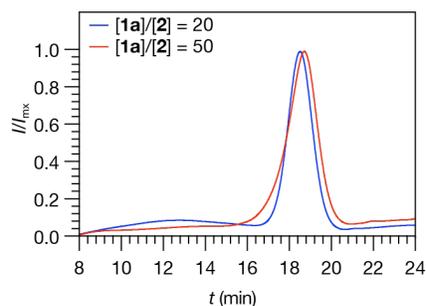

**Extended Data Figure ED1 | *Suzuki* catalyst transfer polymerization with monomer 1a.** SEC traces of *poly*-**1a** samples obtained from $[M]_0/[I]_0 = 10$ and $[M]_0/[I]_0 = 50$ loadings.

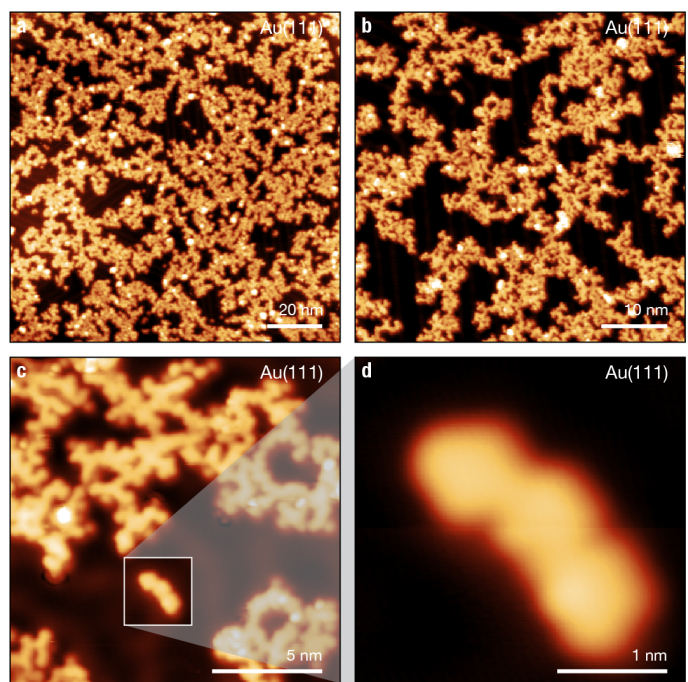

**Extended Data Figure ED2 | STM characterization of MAD transferred samples of solution-cyclodehydrogenated 9-AGNR-4b derived from *poly*-1b. a–b,** Large-area topographic STM images (constant current) of MAD transferred annealed samples of 9-AGNR-**4b** cyclodehydrogenated in solution using *Scholl* or DDQ/TfOH reaction conditions on Au(111) ($V_s$ = 200 mV, $I_t$ = 20 pA). **c–d,** Representative topographic STM images (constant current) of MAD transferred samples of solution cyclodehydrogenated 9-AGNR-**4b** on Au(111) showing a high concentration of structural defects and an internal bonding that cannot be mapped onto the structure of the 9-AGNR backbone ($V_s$ = 200 mV, $I_t$ = 20 pA).

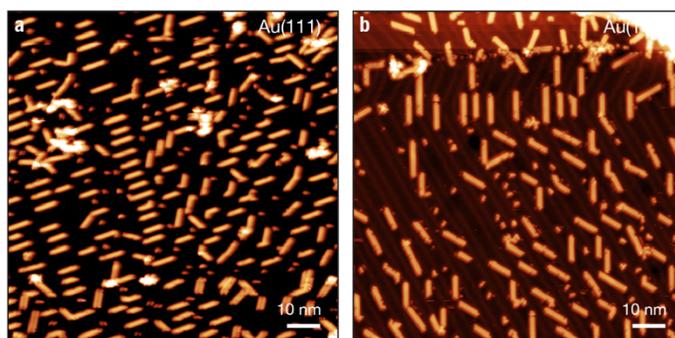

**Extended Data Figure ED3 | Hybrid surface assisted synthesis of 9-AGNRs-4a from polymer precursors. a–b,** Large-area topographic STM images (constant current) of a MAD transferred annealed sample of *poly*-**4a** ($[M]_0/[I]_0 = 20$) on Au(111) ($V_s = 100$ mV, $I_t = 20$ pA).

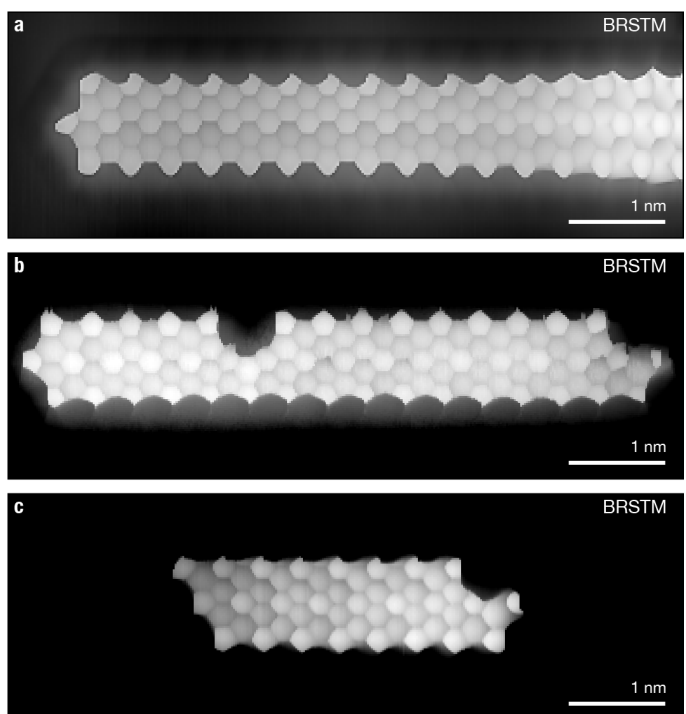

**Extended Data Figure ED4 | Hybrid surface assisted synthesis of 9-AGNRs from polymer precursors. a,** BRSTM image (constant height) of a defective 9-AGNR-**4a** grown on Au(111) from *poly*-**4a** ($[M]_0/[I]_0 = 20$) using the hybrid SCTP/surface-assisted cyclodehydrogenation approach ($V_s = 10$ mV, $V_{ac} = 10$ mV, $f = 455$ Hz, CO-functionalized tip). The end-group transferred during the termination step (right edge in **a**) is fused to the edge of a second ribbon on the surface. **b,** BRSTM image (constant height) of a defective 9-AGNR-**4a** grown on Au(111) from *poly*-**4a** ($[M]_0/[I]_0 = 20$) using the hybrid SCTP/surface-assisted cyclodehydrogenation approach ($V_s = 10$ mV, $V_{ac} = 10$ mV, $f = 455$ Hz, CO-functionalized tip). The edge

defect results from the cleavage of a phenyl ring during the cyclodehydrogenation on the surface. **c,** BRSTM image (constant height) of a defective 9-AGNR-**4a** grown on Au(111) from *poly*-**4a** ([M]$_0$/[I]$_0$ = 20) using the hybrid SCTP/surface-assisted cyclodehydrogenation approach ($V_s$ = 10 mV, $V_{ac}$ = 10 mV, $f$ = 455 Hz, CO-functionalized tip). The biphenyl end group transferred during the initiation step has been cleaved on the surface. All STM experiments performed at $T$ = 4 K.

**Supplementary Information**

Supplementary Information contains detailed synthetic procedures and characterization of precursor **1a,b**, **2**, and polymers *poly*-**1a,b** and *poly*-**4a,b**.